\documentclass[12pt]{article}
\usepackage[]{graphicx}
\usepackage[]{color}

\usepackage{alltt}
\usepackage[T1]{fontenc}
\usepackage[latin9]{inputenc}
\usepackage{geometry}
\usepackage{array}
\usepackage{amsmath}
\usepackage{graphicx}
\usepackage[authoryear]{natbib}
\usepackage[unicode=true]{hyperref}

\makeatletter

\providecommand{\tabularnewline}{\\}

\newcommand{\lyxaddress}[1]{
\par {\raggedright #1
\vspace{1.4em}
\noindent\par}
}

\usepackage{times}
\usepackage{graphics,natbib}

\newcommand{\matern}{Mat\'{e}rn }

\makeatother

\begin{document}

\title{Statistically-estimated tree composition for the northeastern United
States \\
at the time of Euro-American settlement}

\author{Christopher J. Paciorek\textsuperscript{1,{*}}, Simon J. Goring\textsuperscript{2,\&},
Andrew L. Thurman\textsuperscript{3,\&}, \\
Charles V. Cogbill\textsuperscript{4}, John W. Williams\textsuperscript{2,3},
David J. Mladenoff\textsuperscript{6}, \\
Jody A. Peters\textsuperscript{7}, Jun Zhu\textsuperscript{8}, and
Jason S. McLachlan\textsuperscript{7}}

\maketitle

\lyxaddress{\textsuperscript{1}Department of Statistics, University of California,
Berkeley, California, USA}

\lyxaddress{\textsuperscript{2}Department of Geography, University of Wisconsin,
Madison, Wisconsin, USA}

\lyxaddress{\textsuperscript{3}VA Office of Rural Health, Veterans Rural Health
Resource Center, Iowa City VAMC, Iowa City, Iowa, USA}

\lyxaddress{\textsuperscript{4}Harvard Forest, Harvard University, Petersham,
Massachusetts, USA}

\lyxaddress{\textsuperscript{5}Center for Climatic Research, University of Wisconsin,
Madison, Wisconsin, USA}

\lyxaddress{\textsuperscript{6}Department of Forest and Wildlife Ecology, University
of Wisconsin, Madison, Wisconsin, USA}

\lyxaddress{\textsuperscript{7}Department of Biological Sciences, University
of Notre Dame, Notre Dame, Indiana, USA}

\lyxaddress{\textsuperscript{8}Department of Statistics, University of Wisconsin,
Madison, Wisconsin, USA}

\lyxaddress{\textsuperscript{{*}}Corresponding author; E-mail: paciorek@stat.berkeley.edu}

\lyxaddress{\textsuperscript{\&}These authors contributed equally to this work.
\newpage}

\begin{abstract}
We present a gridded 8 km-resolution data product of the estimated
composition of tree taxa at the time of Euro-American settlement of
the northeastern United States and the statistical methodology used
to produce the product from trees recorded by land surveyors. Composition
is defined as the proportion of stems larger than approximately 20
cm diameter at breast height for 22 tree taxa, generally at the genus
level. The data come from settlement-era public survey records that
are transcribed and then aggregated spatially, giving count data.
The domain is divided into two regions, eastern (Maine to Ohio) and
midwestern (Indiana to Minnesota). Public Land Survey point data in
the midwestern region (ca. 0.8-km resolution) are aggregated to a
regular 8 km grid, while data in the eastern region, from Town Proprietor
Surveys, are aggregated at the township level in irregularly-shaped
local administrative units. The product is based on a Bayesian statistical
model fit to the count data that estimates composition on a regular
8 km grid across the entire domain. The statistical model is designed
to handle data from both the regular grid and the irregularly-shaped
townships and allows us to estimate composition at locations with
no data and to smooth over noise caused by limited counts in locations
with data. Critically, the model also allows us to quantify uncertainty
in our composition estimates, making the product suitable for applications
employing data assimilation. We expect this data product to be useful
for understanding the state of vegetation in the northeastern United
States prior to large-scale Euro-American settlement. In addition
to specific regional questions, the data product can also serve as
a baseline against which to investigate how forests and ecosystems
change after intensive settlement. The data product is being made
available at the NIS data portal as version 1.0.

\end{abstract}

\section{Introduction}

Historical datasets provide critical context to understand forest
ecology. They allow researchers to define `baseline' conditions for
conservation management, to understand ecosystem processes at decadal
and centennial scales, to track forest responses to shifting climates,
and, particularly in regions with widespread land use change, to understand
the extent to which forests after conversion and regeneration differ
from the original forest cover.

Euro-American settlement and subsequent land use change occurred in
a time-transient fashion across North America and were accompanied
by land surveys needed to demarcate land for land tenure and use.
Various systems were used by surveyors to locate legal boundary markers,
usually by recording and marking trees adjacent to survey markers.
These data provide vegetation information that can be mapped and used
quantitatively to represent the period of settlement. Early surveys
(from 1620 until 1825) in the northeastern United States provide spatially-aggregated
data at the township level \citep{Cogb:etal:2002,thompson2013four},
with typical township size on the order of 200 $\mbox{km}^{2}$ and
no information about the locations of individual trees; we refer to
these as the Town Proprietor Survey (TPS). Later surveys after the
establishment of the U.S. Public Land Survey System (PLS) by the General
Land Office (GLO) provide point-level data along a regular grid, with
one-half mile (800 m) spacing, for Ohio and westward during the period
1785 to 1907 \citep{bourdo1956review,pattison1957beginnings,schulte2001original,goring2015composition}.
At each point 2-4 trees were identified, and the common name, diameter
at breast height, and distance and bearing from the point were recorded.
Survey instructions during the PLS varied through time and by point
type. Accounting for this variation requires data screening to maximize
consistency among points and the application of spatially-varying
correction factors \citet{goring2015composition} to accurately assess
tree stem density, basal area and biomass from the early settlement
records, but the impact on composition estimates is limited \citep{liu2011broadscale}.
Surveyors sometimes used ambiguous common names, which requires matching
to scientific names and standardization \citep{mladenoff2002narrowing,goring2015composition}.

Logging, agriculture, and land abandonment have left an indelible
mark on forests in the northeastern United States \citep{foster1998land,rhemtulla2009legacies,thompson2013four,goring2015composition}.
However most studies have assessed these effects in individual states
or smaller domains \citep{friedman2005regional,rhemtulla2009historical}
and with various spatial resolutions, from townships (36 square miles)
to forest zones of hundreds or thousands of square miles. \citet{goring2015composition}
provide a new dataset of forest composition, biomass, and stem density
based on PLS data for the upper Midwest that is resolved to an 8 km
by 8 km grid cell scale, providing broad spatial coverage at a spatial
scale that can be compared to modern forests using Forest Inventory
and Analysis products \citep{gray2012forest}. Combined with additional,
coarsely-sampled PLS data from Illinois and Indiana, newly-digitized
data from southern Michigan, and with the TPS data, this gives us
raw data for much of the northeastern United States. However, there
are several limitations of using the raw data that can be alleviated
by the use of a statistical model to develop a statistically-estimated
data product. First, the PLS and TPS data only provide estimates of
within-cell variance that do not account for information from nearby
locations. Second, there are data gaps: the available digitized data
from Illinois and Indiana represent a small fraction of those states,
and missing townships are common in the TPS data. Third, the TPS and
PLS data have fundamentally different sampling design and spatial
resolution. Our statistical model allows us to provide a spatially-complete
data product of settlement-era tree composition for a common 8 km
grid with uncertainty across the northeastern U.S.

In Section \ref{sec:Data} we describe the data sources, while Section
\ref{sec:Statistical-model} describes our statistical models. In
Section \ref{sec:Model-comparison} we quantitatively compare competing
statistical specifications, and in Section \ref{sec:Data-product}
we describe the final data product. In Section \ref{sec:Discussion}
we discuss the uncertainties estimated by and the
limitations of the statistical model, and we list related data products
under development.

\section{Data\label{sec:Data}}

\begin{figure}
\label{fig:domain}\includegraphics[scale=1.3]{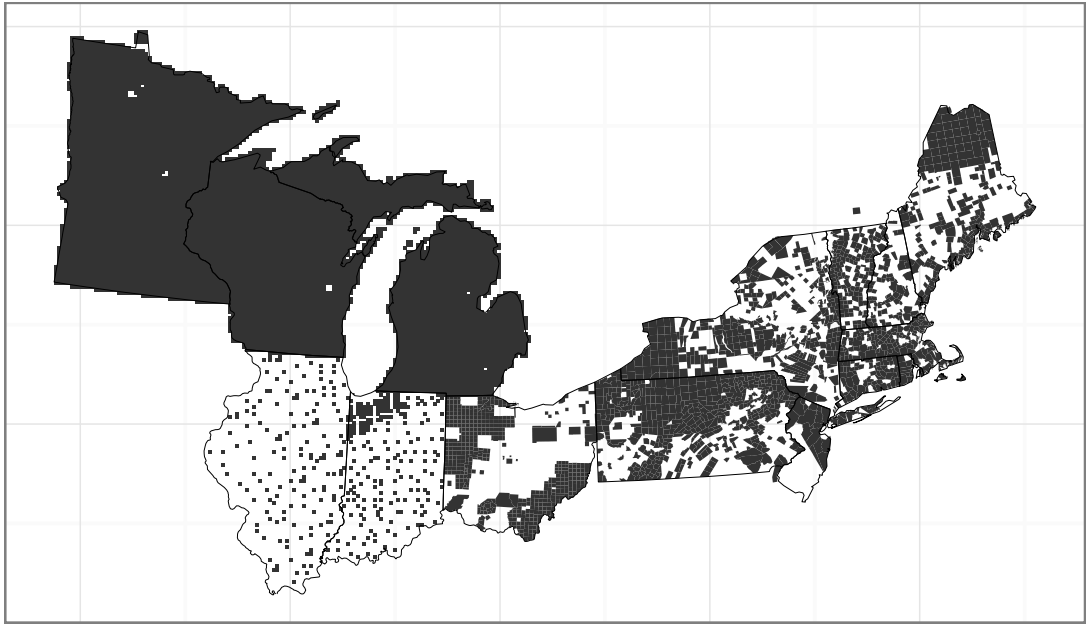}

\caption{Spatial domain of the northeastern United States, with locations with
data shown in gray. Locations are grid cells in midwestern portion
and townships in eastern portion. In addition to locations without
data being indicated in white, grid cells completely covered in water
are white (e.g., a few locations in the northwestern portion of the
domain in the states of Minnesota and Wisconsin).\label{fig:Spatial-domain}}
\end{figure}

The raw data were obtained from land division survey records collated
and digitized from across the northeastern U.S. by a number of researchers
(Fig. \ref{fig:Spatial-domain}). For the states of Minnesota, Wisconsin,
Illinois, Indiana, and Michigan (the midwestern subdomain), digitized
data are available at PLS survey point locations and have been aggregated
to a regular 8 km grid in the Albers projection. (Note that for Indiana
and Illinois, at the moment trees are associated with township centroids
and then assigned to 8 km grid cells based on the centroid, but in
the near future we will have point locations available for each tree.)
For the states of Ohio, Pennsylvania, New Jersey, New York and the
six New England states (the eastern subdomain), data are aggregated
at the township level. We make predictions for all
of the states listed above; these constitute our core domain. There
are also data from a single township in Quebec and a single township
in northern Delaware; these data help inform predictions
in nearby locations within our core domain, but predictions are not
made for Quebec or Delaware. Digitization of PLS data in Minnesota,
Wisconsin and Michigan is essentially complete, with PLS data for
nearly all 8 km grid cells, but data in Illinois and Indiana represent
a sample of the full set of grid cells, with survey record transcription
ongoing. Data for the eastern states are available for a subset of
the full set of townships covering the domain; the TPS data for some
townships were lost, incomplete, or have not been located \citep{Cogb:etal:2002}. 

Note that surveys occurred over a period of more than 200 years as
European colonists (before U.S. independence) and the United States
settled what is now the northeastern and midwestern United States.
Our estimates are for the period of settlement represented by the
survey data and therefore are time-transgressive; they do not represent
any single point in time across the domain, but rather the state of
the landscape at the time just prior to widespread Euro-American settlement
and land use \citep{Whit:1996,Cogb:etal:2002}. These forest composition
datasets do include the effects of Native American land use and early
Euro-American settlement activities \citep[e.g.,][]{Blac:etal:2006},
but it is likely that the imprint of this earlier land use is highly
concentrated rather than spatially extensive \citep{munoz2014defining}.

Extensive details on the upper Midwest (Minnesota, Wisconsin, Michigan)
data and processing steps are available \cite{goring2015composition};
key elements include the use of only corner points, the use of only
the two closest trees at each corner point, spatially-varying correction
factors for sampling effort, and a standardized taxonomy table. The
lower Midwest (Illinois, Indiana) data were purchased from the Indiana
State Archives (Indiana) and Hubtack Document Resources (hubtack.com;
Illinois) and processed using similar steps as for the upper Midwest
data. Digitization of the Illinois and Indiana data is still underway,
so many grid cells contained no data at the time the statistical model
was fit. Note that the number of trees per grid cell varies depending
on the number of survey points in a cell, with an average of 124 trees
per cell. The gridded data at the 8 km resolution for the midwest
subdomain are available through the NIS data portal \citep{Gori:etal:data:2016}.
The TPS data were compiled by C.V. Cogbill from a myriad of archival
sources representing land division surveys conducted in connection
with local settlement and are available through the
NIS data portal \citep{Cogb:dataNE:2016,Cogb:dataOH:2016}. 

The aggregation into taxonomic groups is primarily at the genus level
but is at the species level in some cases of monospecific genera.
We model the following 22 taxa plus an ``other hardwood'' category:
Atlantic white cedar (\emph{Chamaecyparis thyoides}), Ash (\emph{Fraxinus
spp.}), Basswood (\emph{Tilia americana}), Beech (\emph{Fagus grandifolia)},
Birch (\emph{Betula spp.}), Black gum/sweet gum (\emph{Nyssa sylvatica}
and \emph{Liquidambar styraciflua}), Cedar/juniper (\emph{Juniperus
virginiana} and \emph{Thuja occidentalis}), Cherry (\emph{Prunus spp.}),
Chestnut (\emph{Castanea dentata}), Dogwood (\emph{Cornus spp.}),
Elm (\emph{Ulmus spp.}), Fir (\emph{Abies balsamea}), Hemlock (\emph{Tsuga
canadensis}), Hickory (\emph{Carya spp.}), Ironwood (\emph{Carpinus
caroliniana} and \emph{Ostrya virginiana}), Maple (\emph{Acer spp.}),
Oak (\emph{Quercus spp.}), Pine (\emph{Pinus spp.}), Poplar/tulip
poplar (\emph{Populus spp.}~and \emph{Liriodendron tulipifera}),
Spruce (\emph{Picea spp.}), Tamarack (\emph{Larix laricina}), and
Walnut (\emph{Juglans spp.}). Note that in several cases (black gum/sweet
gum, ironwood, poplar/tulip poplar, cedar/juniper), because of ambiguity
in the common tree names used by surveyors, a group represents trees
from different genera or even families. For the midwestern subdomain
we do not fit statistical models for Atlantic white cedar and chestnut
as these species have 0 and 7 trees present, respectively. The taxa
grouped into the other hardwood category are those for which fewer
than roughly 2000 trees were present in the dataset; however, we include
Atlantic white cedar explicitly despite it only having 336 trees in
the dataset because of specific ecological interest in Atlantic white
cedar wetlands. 

Diameters are only recorded in the PLS data. Although surveyors avoided
using small trees, there was no consistent lower diameter limit. The
PLS data generally represent trees greater than 8 inches (\textasciitilde{}20
cm) diameter at breast height (dbh), but with some trees as small
as 1 inch dbh (smaller trees were much more common in far northern
Minnesota). TPS data have no information about dbh, but the trees
were large enough to blaze and are presumed to be relatively large
trees useful for marking property boundaries.

There are approximately 860,000 trees in the midwestern subdomain
and 420,000 trees in the eastern subdomain. In the midwestern subdomain,
oak is the most common taxon and pine the second most common, while
in the eastern subdomain oak is the most common and beech the second
most common.

Our domain is a rectangle covering all of the states using a metric
Albers (Great Lakes and St. Lawrence) projection (PROJ4: EPSG:3175),
with the rectangle split into 8 km cells, arranged in a 296 by 180
grid of cells, with the centroid of the cell in the southwest corner
located at (-71000 m, 58000 m). For the midwestern subdomain we use
the western-most 146 by 180 grid of cells when fitting
the statistical models. For the eastern subdomain we use the eastern-most
180 by 180 grid of cells and then omit 23 rows of cells in the north
and 17 rows of cells in the south, as these grid cells are outside
of the states containing data.

\section{Statistical model\label{sec:Statistical-model}}

We fit a Bayesian statistical model to the data, with two primary
goals:
\begin{enumerate}
\item To estimate composition on a regular grid across the entire domain,
filling gaps where no data are available, and
\item To quantify uncertainty in composition at all locations. Even in grid
cells and townships with data, we wish to quantify uncertainty because
the empirical proportions represent estimates of the true proportions
that could be calculated using the full population of all the trees
in a grid cell or township.
\end{enumerate}
At a high level, the Bayesian statistical model estimates
composition across the domain, even in locations with sparse or no
data, by combining the raw composition data with the assumption that
composition varies in a smooth spatial fashion across the domain.
The information in the data is quantified by the data model, also
known as the likelihood. The assumption of smoothness is built into
the model by representing the true unknown spatially-varying composition
using a statistical spatial process representation that induces smoothing
of estimates across nearby locations. This spatial process representation
is a form of prior distribution and is a function of model parameters
called hyperparameters that determine the correlation structure of
the process and are also estimated based on the data.

The result of fitting the Bayesian model via Markov chain Monte Carlo
(MCMC) is a set of representative samples from the posterior distribution
for the composition in the 23 taxonomic groupings at each of the grid
cells. These samples are the data product (described further in the
Section \ref{sec:Data-product}) and can be used in subsequent analyses.
The mean and standard deviation of the samples for each pair of cell
and taxon represent our best estimate (i.e., prediction) of composition
and a Bayesian ``standard error'' quantifying the uncertainty in
the estimate. 

In the remainder of this section we provide the technical
specification of the model and of the computations involved in fitting
the model.

\subsection{Data model}

We start by describing the basic model for those states for which
we have raw data on the 8 km grid, and in Section \ref{sub:Model-for-township}
we describe the extension of the model to accommodate data aggregated
at the township level.

The statistical model treats the observations as coming from a multinomial
distribution with a (latent) vector of proportions for each grid cell,
\[
y_{i}\sim\mbox{Multi}(n_{i},\theta(s_{i})),
\]
where $y_{i}$ is the vector of counts for the $P$ taxa at the $i$th
cell, $n_{i}$ is the number of trees counted in the cell, and $\theta(s_{i})$
is the vector of unknown proportions for those taxa at that cell.
Note that we use a standard multinomial distribution without overdispersion,
because the set of trees in the dataset is roughly uniformly sampled
across the cells or townships \citep{goring2015composition}.

The proportions, $\theta_{p}(s_{i}),\, p=1,\ldots,P$, are modeled
spatially by a set of $P$ Gaussian spatial processes, one per taxon,
$\alpha_{p}(s_{i}),\, p=1,\ldots,P$. This collection of processes
defines a multivariate spatial process for composition. The $\alpha_{p}(s)$
processes are defined on the 8 km grid, $\alpha_{p}=\{\alpha_{p}(s_{1}),\ldots,\alpha_{p}(s_{m})\}$
for the $m$ grid cells. In Section \ref{sub:Latent-Variable-Model}
we introduce a multinomial probit model that relates the $\alpha_{p}(s)$
processes to the proportion processes, $\theta_{p}(s)$, via the introduction
of latent variables, with an implicit sum-to-one constraint, $\sum_{p=1}^{P}\theta_{p}(s)=1$.
A multinomial probit model is similar to logistic
regression, used for modeling a binary outcome based on an underlying
probability the outcome will occur, but generalizes to modeling a
categorical variable based on probabilities for each category.

The critical component of the statistical model is the representation
of $\alpha_{p}(s)$ as a spatial process. A spatial
process is a statistical representation that models spatially-correlated
values. It provides a prior structure that serves to smooth across
noise in the observations and allows for prediction at locations based
on information from nearby locations, including interpolation to locations
with no data. Apart from the sum-to-one constraint, the taxa are considered
to be independent in the prior. We did not want to impose any structure
that ties the different taxa together, as any correlation will likely
vary across space.

In the next section, we consider two spatial models to define the
structure of the $\alpha_{p}(s)$ processes, a standard conditional
autoregressive model \citep{Bane:etal:2004} and a Gaussian Markov
random field (MRF) approximation to a Gaussian process with \matern
covariance \citep{Lind:etal:2011}. These models are
specific statistical formulations of spatial processes that represent
spatial correlation by defining neighborhoods around each location
that are used to help inform predictions at the location.

\subsection{Spatial process models}

MRF models represent the neighborhood information
by working directly with the precision matrix (the
inverse of the covariance matrix) of the values of the spatial process,
so calculation of the prior density of $\alpha_{p}$ is computationally
simple \citep{Rue:Held:2005}. However, in situations
in which the likelihood is not normal, such
as our multinomial likelihood, it can be difficult to set up effective
MCMC algorithms that are able to move in the high-dimensional space
of $\alpha_{p}$. The latent variable representation of Section \ref{sub:Latent-Variable-Model}
helps to alleviate this problem. Next we describe two alternative
spatial models that we considered; in Section \ref{sec:Model-comparison},
we evaluate the models on held-out data to choose
between the two.

\paragraph{Standard conditional autoregressive models}

Our first model is a standard conditional autoregressive (CAR) model;
technical details can be found in \citep{Bane:etal:2004}. We use
a standard form of this model that treats the four
cardinal neighbors of each grid cell as the neighbors of the grid
cell. The corresponding precision matrix has diagonal elements, $Q_{ii}$,
equal to the number of neighbors for the $i$th area (i.e., four except
for cells on the boundary of the domain), while $Q_{ik}=-1$ (the
negative of a weight of one) when areas $i$ and $k$ are neighbors
and $Q_{ik}=0$ when they are not. This gives the following model
for the values of $\alpha_{p}(s_{i})$ collected as a vector across
all of the grid cells, $i=1,\ldots,m$: 
\[
\alpha_{p}\sim\mbox{N}(0,\sigma_{p}^{2}Q^{-}).
\]
The use of the generalized inverse notation indicates that $Q$ is
not full-rank, but is of rank $m-1$; this gives an improper prior
on an implicit overall mean for the process values. Note
that we specify an explicit mean of zero because a non-zero mean would
not be identifiable in light of the implicit mean. This specification
is called an \textit{intrinsic conditional autoregression (ICAR)}
and we can write $Q=D-C$ where $C$ is the $m\times m$ adjacency
matrix defining the neighborhood relation of the locations; that is
$C{}_{ik}=1$ if locations $i$ and $k$ are neighbors and zero otherwise.
The matrix $D$ is an $m\times m$ diagonal matrix containing the
row sums of matrix $C$ as the diagonal entries, $D{}_{ii}={\displaystyle \sum_{k=1}^{m}C{}_{ik}}.$

We refer to this as the \emph{CAR model}.

\paragraph{Gaussian process approximation}

Gaussian processes (GP) are also standard models for spatial processes \citep{Bane:etal:2004}.
GP models are computationally challenging for large datasets because
of computational manipulations involving large covariance
matrices. Given this, \cite{Lind:etal:2011} proposed a new framework
for using Gaussian MRFs (GMRFs) as approximations to GPs, based on
the use of stochastic partial differential equations (SPDEs).

Gaussian processes are generally constructed using
one of a number of correlation functions that define how the strength
of correlation between the values of the process at two locations
decays as a function of the distance between the locations. We consider
Gaussian processes in the commonly-used \matern
class, using the following parameterization of the \matern correlation
function, 
\begin{equation}
R(d)=\frac{1}{\Gamma(\nu)2^{\nu-1}}\left(\frac{2\sqrt{\nu}d}{\rho}\right)^{\nu}\mathcal{K}_{\nu}\left(\frac{2\sqrt{\nu}d}{\rho}\right),\nonumber
\end{equation}
where $d$ is Euclidean distance, $\rho$ is the spatial range parameter,
and $\mathcal{K}_{\nu}(\cdot)$ is the modified Bessel function of
the second kind, whose order is the smoothness (differentiability)
parameter, $\nu>0$. $\nu=0.5$ gives the exponential covariance.
For any pair of locations, $R(d)$ defines the correlation of the
process, (i.e., $\alpha_{p}(s)$ in our context), as a function of
the distance between the locations. Considering all pairs of locations,
this defines a correlation matrix for all locations of interest. 

The approach of \citet{Lind:etal:2011} allows us to consider MRF
approximations to the \matern-based GP for $\nu=1$ and $\nu=2$.
Our second spatial model is this Lindgren approximation for \matern-based
GPs with $\nu=1$. To implement this Lindgren model, one modifies
the $Q$ matrix defined previously as follows based
on the technical specification of the precision matrix provided in
\citet{Lind:etal:2011}. Let $a=4+\frac{1}{\rho^{2}}$. The diagonal
elements of $Q$ are $4+a^{2}$. The entries corresponding to cardinal
neighbors are $-2a$. Those for diagonal neighbors are $2$, and those
for 2nd-order cardinal neighbors are $1$. This extends the neighborhood
structure relative to the CAR model and parameterizes it as a function
of $\rho$.

The primary difference between the CAR and Lindgren models is that
the Lindgren model provides an additional degree of freedom by estimating
$\rho$. In particular $\rho$ allows us to estimate the locality
of the spatial smoothing. As $\rho$ decreases, the
model uses increasingly localized data to estimate the compositional
proportions at a given location, effectively averaging the empirical
proportions over smaller neighborhoods. In general, the \cite{Lind:etal:2011}
model will generally provide for a smoother estimate than the CAR
model \citep{Paci:2013}. 

To ensure that the $\sigma^{2}$ parameter is mathematically equivalent
between the two models, we reparameterize, producing our second model:
\[
\alpha_{p}\sim\mbox{N}\left(\mu_{p},\sigma_{p}^{2}\cdot\frac{4\pi}{\rho_{p}^{2}}Q(\rho_{p})^{-1}\right)
\]

We refer to this model as the \emph{SPDE model}.

\subsection{Prior Distributions}

\noindent The ICAR specification contains a set of hyperparameters
$\{\sigma_{p}^{2}\},$ $p=1,\ldots P$. Following \cite{Gelm:2006}
we use a uniform distribution on each $\sigma_{p}$ parameter, with
upper bound of 1000. For the SPDE model we also have hyperparameters
$\{\mu_{p}\}$, which we give flat, non-informative priors (truncated
at $\pm10$), and $\{\rho_{p}\}$, which we give uniform priors on
the interval $(0.1,\exp(5))$. These various hyperparameters
are unknown parameters that control the spatial structure of the two
spatial models and are estimated from both the data and the prior
distributions just specified based on the Bayesian approach.

\subsection{Latent Variable Model\label{sub:Latent-Variable-Model}}

It is well-known that devising an effective MCMC algorithm for models
with latent Gaussian process(es) and a non-Gaussian likelihood is
difficult \citep{Rue:Held:2005,Chri:etal:2006,Tan:Nott:2013}. To
develop an algorithm, we make use of a latent variable representation
for the multinomial probit model \citep{McCu:Ross:1994}. The representation
introduces latent variables that allow one to develop an MCMC sampling
strategy that takes advantage of closed-form full conditional distributions
(so-called Gibbs sampling steps) for $\alpha_{p}$.

Suppose that compositional counts are available at a number of locations.
At location $i$, a sample of size $n_{i}$ observations is collected,
and each observation (i.e., each tree) can be classified into $P$
distinct categories. For a given tree $j$ at location $i$, let $Y_{ij}$
denote the response variable indicating the category. Let $Y_{ij}$
be associated with $P$ latent variables $W_{ij1},...,W_{ijP}$ such
that $Y_{ij}$ = $p$ if and only if $W_{ijp}={\displaystyle \max_{p'}\big\{ W_{ijp'}\big\}}$;
in other words, the maximum of the set of latent variables $\{W_{ijp}\}{\displaystyle _{p=1}^{P}}$
determines the category of observation $j$ at location $i$. The
final piece of the latent variable representation is the relationship
between the $W$ variables and the $\alpha_{p}(s)$ processes. We
have that

\noindent 
\[
W_{ijp}\sim\mbox{N}(\alpha_{p}(s_{i}),1)
\]
independently for all of the $W_{ijp}$ values. Consider the following
example with two locations that are neighbors and $P=2$ categories.
Each tree $j$ at location $i$ is associated with two variables $W_{ij1}$
and $W_{ij2}$, governed by the latent variables $\alpha_{1}(s_{i})$
and $\alpha_{2}(s_{i})$, respectively. Suppose that $\alpha_{1}(s_{i})>\alpha_{2}(s_{i})$
for a given location $i$. Then this model implies that any tree $j$
is more likely to be labeled 1 than 2 at location $i$. The difference
between $\alpha_{1}(s_{i})$ and $\alpha_{2}(s_{i})$ explains the
\textit{difference} in probability of \textit{categories} 1 and 2
at location $i$, and the similarity between $\alpha_{p}(s_{1})$
and $\alpha_{p}(s_{2})$ explains the \textit{correlation} between
the probabilities at \textit{locations} 1 and 2 for category $p$.

\noindent

\noindent

\subsection{Model for township data\label{sub:Model-for-township}}

We developed an extension of the model described in previous sections
to account for data at a different aggregation than our core 8 km
grid. This extension introduces a new set of latent variables, one
per tree, that indicate the grid cells in which the trees are located
and that can be sampled within the MCMC as additional unknown parameters.
Specifically, $c_{tj}$ is the latent ``membership'' variable for
tree $j$ in township $t$, $t=1,\ldots,T$. The prior for $c_{tj}$
is a discrete distribution that puts mass, $\psi_{ti}$, $i=1,\ldots,m$,
proportional to the areal overlap between the township in which the
tree is located and the $m$ grid cells, giving 
\[
c_{tj}\sim\mbox{Multinom}(1,\{\psi_{t1},\ldots,\psi_{tm}\}),
\]
independently across all trees. Because the townships overlap a limited
number of grid cells, most of the $\psi_{t1},\ldots,\psi_{tm}$ values
are zero.

Using the latent variable representation, we have that $W_{tjp}\sim\mbox{N}(\alpha_{p}(s_{c_{tj}}),1)$
for tree $j$ in township $t$. In updating the other parameters in
the model during the MCMC (specifically the $\alpha$ values), we
condition on the current values, $\{c_{tj}\}$, which provides a ``soft''
(i.e., probabilistic) assignment of trees to grid cells that respects
both the known township in which the trees occurred and the uncertainty
in which grid cells the trees occurred.

Note that this prior represents the location of each tree in a township
as being independent of the other trees; this is somewhat unrealistic
because it does not represent our knowledge that the trees in a township
would be distributed somewhat regularly across the area of the township
because the witness trees were used to indicate property boundaries.

\subsection{Computation}

The \cite{McCu:Ross:1994} representation is convenient for MCMC sampling,
particularly in this high-dimensional spatial context, as it allows
us to draw from the posterior conditional distributions of the $W_{ijp}$
variables (these distributions are truncated normal) in closed form
and to draw the entire vector of latent process values for each taxon,
$\alpha_{p}$, as a single sample that respects the spatial dependence
structure for each taxon.

While the latent variable representation provides great advantages
in the MCMC sampling for each $\alpha_{p}$ compared to joint Metropolis
updates or updating each location individually, there is still strong
dependence between the hyperparameters, $\{\sigma_{p}^{2},\mu_{p},\rho_{p}\}$
and the latent process values (as well as between the latent process
values and the latent $W_{ijp}$ variables). To address the first,
we developed a ``cross-level'' joint updating strategy for the CAR
model in which we propose $\phi_{p}=\sigma_{p},p=1,\ldots,P,$ (and
for the SPDE model, $\phi_{p}\in\{\mu_{p},(\sigma_{p},\rho_{p})\}$)
via a Metropolis-style random walk and then given the proposed value,
$\phi_{p}^{*}$, propose $\alpha_{p}$ from its full conditional distribution
given $\phi_{p}^{*}$ and the latent $W_{p}$ variables, where $W_{p}$
is the vector of all $W_{ijp}$ values for taxon $p$: $W_{p}=\{W_{ijp}\},i=1,\ldots,m;j=1,\ldots,n_{i}$.
This is equivalent to sampling from the marginalized (with respect
to $\alpha_{p}$) distribution of $\phi_{p}$ conditional on $W_{p}$.
For these various joint samples of hyperparameters and $\alpha_{p}$,
we use adaptive Metropolis sampling \citep{Shab:Well:2011}.

The full description of the MCMC sampling steps is provided in the
Appendix. In addition, in the latent variable representation, $\theta_{p}(s)$
never appears explicitly and cannot be calculated in close form. Instead
we use Monte Carlo integration over $W_{ijp},\, p=1,\ldots,P$ to
estimate $\theta_{p}(s_{i})$, also described in the Appendix. 

The model is implemented in R \citep{R:2014} with core computational
calculations coded in C++ using the \emph{Rcpp} package \citep{Edde:Fran:2011}.
We also make extensive use of sparse matrix representations and algorithms,
using the \emph{spam} package in R \citep{Furr:Sain:2010}. All code
is available on Github, including pre- and post-processing code, at\\\url{https://github.com/PalEON-Project/composition}.

\section{Model comparison\label{sec:Model-comparison}}

\subsection{Design}

We compared the CAR and SPDE models by holding out
data from the fitting process and assessing the fit of the model
on the held-out data. We used two experiments\textcolor{red} with
held-out data:
\begin{enumerate}
\item The first experiment used a subregion containing most of Minnesota
and a small amount of western Wisconsin, defined to be the cells whose
x-coordinate was less than 300,000 m (this defines a north-south line
that approximately goes through Duluth, Minnesota) and hereafter referred
to as the ``Minnesota subregion''. We chose this subregion for evaluation
because of its high data density, allowing us to experiment with the
effects of increasing data sparsity on model performance. We held
out all data from 95\% of the cells in this Minnesota subregion, with
cells selected at random. This was meant to assess the ability of
the model to interpolate from a sparse set of cells/townships and
mimics the limited data in Illinois and Indiana.
\item We held out 5\% of the trees from all of the trees in the dataset
for the midwestern subdomain (leaving aside the held-out Minnesota
subregion cells). This was meant to assess the ability of the model
to estimate the composition in cells in which data were available. 
\end{enumerate}
Finally, in a separate sensitivity analysis we instead left out 80\%
of the cells in Minnesota subregion at random. This variation on the
first experiment above was meant to indicate whether our model comparison
conclusions would be robust as the digitization process for Illinois
and Indiana progresses and provides us with increasingly dense data. 

There has been extensive work in the statistical literature on good
metrics to use to compare the predictive ability of models; these
metrics are referred to as scoring rules. A general conclusion from
this work is that predictive distributions should maximize sharpness
subject to calibration. That is, the predictive distribution should
be as narrow as possible while being calibrated such that the observations
are consistent with the distribution \citep{Gnei:etal:2007}. When
thinking in terms of prediction intervals as summaries of the predictive
distribution, we seek intervals that are as narrow as possible while
still covering the truth the expected proportion (e.g., 95\% for a
95\% prediction interval) of the time. 

Following the suggestions in \cite{Gnei:etal:2007}, we considered
the following metrics: Brier score, log predictive density, mean square
prediction error, mean absolute error, and coverage and length of
prediction intervals. Further details on each are given below. For
experiment 1, we define $Y_{i}=\{Y_{i1},\ldots,Y_{iP}\}$ as the count
of all trees in held-out cell $i$ and for experiment 2, $Y_{i}$
is the count of held-out individual trees in the cell, while $y_{ijp}$
is an indicator variable taking value either 0 or 1 depending on whether
the $j$th held-out tree in the $i$th cell is of taxon $p$. $\hat{\theta}_{ip}=Y_{ip}/n_{i}$
is the empirical proportion in category $p$ for the $n_{i}$ held-out
trees in cell $i$. We calculated each of the metrics in two ways.
First, we used the posterior mean composition estimates (as an evaluation
of our core predictions), with $\tilde{\theta}_{p}(s)$ being the
posterior mean. Second, we averaged the metric over the posterior
samples (as an evaluation of our full data product, including uncertainty),
taking $\tilde{\theta}_{p}(s)$ to be an individual MCMC sample and
then averaging the metric over all the posterior samples. 
\begin{enumerate}
\item Brier score: \cite{Gnei:etal:2007} suggest this metric, which has
been in use for decades. For multi-category as opposed to binary outcomes,
this takes the form
\[
\frac{1}{n}\sum_{i=1}^{m}\sum_{j=1}^{n_{i}}\sum_{p=1}^{P}(y_{ijp}-\tilde{\theta}_{p}(s_{i}))^{2}
\]
where $n=\sum_{i=1}^{m}n_{i}$ is the total number of held-out trees
for a given experiment and $j$ indexes across held-out trees in cell
$i$. 
\item Log predictive density: This metric takes the log of the probability
density of held-out observations under the fitted model, $Y_{i}\sim\mbox{Multinom}(n_{i},\{\tilde{\theta}_{1}(s_{i}),\ldots,\tilde{\theta}_{P}(s_{i})\})$,
summing on the log scale across all of the held-out data.

While in principle, this metric should be optimal \citep{Krnj:Drap:2014},
it is very sensitive to small predictions near zero \citep{Gnei:etal:2007}.
Even worse, our Monte Carlo estimation of $\theta$ used 10000 samples,
so in some cases $\tilde{\theta}_{p}(s)=0$. When a tree is present
in a cell but its corresponding proportion is 0, this gives a log
density of $-\infty$, preventing use of the metric. As an informal
solution to this we set $\tilde{\theta}_{p}(s)=\frac{1}{100000}$
in such cases, but given these issues we treat the log predictive
density as a secondary measure.

\item (Experiment 1 only) Weighted root mean square prediction error (RMSPE),

\[
\sqrt{\frac{1}{Pn}\sum_{i=1}^{m}\sum_{p=1}^{P}n_{i}(\hat{\theta}_{ip}-\tilde{\theta}_{p}(s))^{2}}
\]
and mean absolute error (MAE)
\[
\frac{1}{Pn}\sum_{i=1}^{m}\sum_{p=1}^{P}n_{i}|\hat{\theta}_{ip}-\tilde{\theta}_{p}(s)|:
\]
These metrics calculate the error of the estimated proportions relative
to the empirical proportions based on the held-out trees, averaging
over cells and taxa. We weight by the number of held-out trees in
each cell to account for the greater variability in the empirical
proportions in locations with few held-out trees. 
\item (Experiment 1 only) Coverage and length of 95\% prediction intervals
for $Y_{ip}$. We considered only cells with at least 50 trees to
focus our assessment on cases where empirical proportions were reasonably
certain and avoid being strongly influenced by predictive inference
for cells where observational variability dominates.
\end{enumerate}
Note that all of the metrics except coverage and interval length can
be applied to individual posterior samples and therefore allow us
to estimate the posterior probability that one model has a lower (better)
value of the metric than the other model by simply calculating the
proportion of samples for which the model has a lower value of the
metric. Also note that in addition to allowing comparison
between models the MAE and RMSPE metrics allow one to assess absolute
performance of each model in predicting composition.

In our initial exploratory fitting, we noticed that the SPDE model
produced boundary effects in the predicted composition near the edges
of the convex hull of the observations. To attempt to alleviate this,
we added a buffer zone with a width of six grid cells around our entire
original domain, but note that the boundary effects were still evident
even after inclusion of the buffer. For the model comparison, we included
this buffer for both the SPDE and CAR models. 

We ran each model for 150,000 iterations. After discarding 25,000
iterations for burn-in, we retained a posterior sample of 250 subsampled
iterations -- we use a subsample instead of the full 125,000 post-burn-in
iterations to reduce post-processing computations and storage needs.

\subsection{Results}

Here we summarize the results of our analyses that inform the choice
between the CAR and SPDE models.

\paragraph{Full cell hold-out experiment}

For Experiment 1 (full cells held out) for cells in the Minnesota
subregion held out of the fitting process, the CAR model outperforms
the SPDE model based on the posterior distribution over the predictive
metric values (Table \ref{tab:score_cell_fivepercent}). For the posterior
mean predictions, the SPDE model appears to outperform the CAR model
to a lesser degree, but we do not have any uncertainty estimates for
this comparison. Coverage and interval lengths are similar between
the two models (Table \ref{tab:coverage_cell_fivepercent}). From
a practical perspective, based on the difference in mean absolute
error, the differences between the models are small (Table \ref{tab:score_cell_fivepercent}).

\begin{table}
\caption{Predictive ability based on several predictive metric criteria for
the CAR and SPDE spatial models when holding out 95\% of entire cells
of data in Minnesota. Smaller values are better.}

\begin{tabular}{|c|c|c|>{\centering}p{3cm}|>{\centering}p{2.5cm}|>{\centering}p{2.5cm}|}
\hline 
 &
\multicolumn{3}{c|}{{\small{}Posterior mean of metric}} &
\multicolumn{2}{c|}{{\small{}Metric of posterior mean predictions}}\tabularnewline
\hline 
\hline 
 &
{\small{}CAR model} &
{\small{}SPDE model} &
{\small{}Posterior Prob. CAR < SPDE} &
{\small{}CAR model} &
{\small{}SPDE model}\tabularnewline
\hline 
{\small{}Brier} &
{\small{}0.819} &
{\small{}0.844} &
{\small{}0.98} &
{\small{}0.738} &
{\small{}0.733}\tabularnewline
\hline 
{\small{}Negative Log Density} &
{\small{}466325} &
{\small{}510383} &
{\small{}1.00} &
{\small{}394003} &
{\small{}394554}\tabularnewline
\hline 
{\small{}Mean Absolute Error} &
{\small{}0.0364} &
{\small{}0.0383} &
{\small{}0.98} &
{\small{}0.0275} &
{\small{}0.0269}\tabularnewline
\hline 
{\small{}Root Mean Square Error} &
{\small{}0.0897} &
{\small{}0.0960} &
{\small{}0.97} &
{\small{}0.0647} &
{\small{}0.0627}\tabularnewline
\hline 
\end{tabular}

\label{tab:score_cell_fivepercent}
\end{table}

\begin{table}
\caption{Coverage and length of prediction intervals for the CAR and SPDE spatial
models when holding out 95\% of entire cells of data in Minnesota.
Coverage values near 0.95 are optimal, while shorter intervals are
better.}

\begin{tabular}{|c|c|c|}
\hline 
 &
{\small{}CAR model} &
{\small{}SPDE model}\tabularnewline
\hline 
{\small{}Coverage} &
{\small{}0.977} &
{\small{}0.978}\tabularnewline
\hline 
{\small{}Mean Interval Length} &
{\small{}0.129} &
{\small{}0.142}\tabularnewline
\hline 
{\small{}Median Interval Length} &
{\small{}0.037} &
{\small{}0.033}\tabularnewline
\hline 
\end{tabular}

\label{tab:coverage_cell_fivepercent}
\end{table}

The results for the variation on Experiment 1 in which the proportion
of cells that are held out decreases from 95\% to 80\% show that the
SPDE model generally outperforms the CAR model, but again differences
from a practical perspective, based on mean absolute error, are limited
(Tables \ref{tab:score_cell_20percent}-\ref{tab:coverage_cell_20percent}).

\begin{table}
\caption{Predictive ability based on several predictive metric criteria for
the CAR and SPDE spatial models when holding out 80\% of entire cells
of data in Minnesota. Smaller values are better.}

\begin{tabular}{|c|c|c|>{\centering}p{3cm}|c|c|}
\hline 
 &
\multicolumn{3}{c|}{{\small{}Posterior mean of score}} &
\multicolumn{2}{c|}{{\small{}Score of posterior mean predictions}}\tabularnewline
\hline 
\hline 
 &
{\small{}CAR model} &
{\small{}SPDE model} &
{\small{}Posterior Prob. CAR < SPDE} &
{\small{}CAR model} &
{\small{}SPDE model}\tabularnewline
\hline 
{\small{}Brier} &
{\small{}0.773} &
{\small{}0.765} &
{\small{}0.10} &
{\small{}0.710} &
{\small{}0.710}\tabularnewline
\hline 
{\small{}Negative Log Density} &
{\small{}355928} &
{\small{}353987} &
{\small{}0.25} &
{\small{}311525} &
{\small{}311902}\tabularnewline
\hline 
{\small{}Mean Absolute Error} &
{\small{}0.0309} &
{\small{}0.0296} &
{\small{}0.10} &
{\small{}0.0226} &
{\small{}0.0223}\tabularnewline
\hline 
{\small{}Root Mean Square Error} &
{\small{}0.0763} &
{\small{}0.0739} &
{\small{}0.02} &
{\small{}0.0533} &
{\small{}0.0530}\tabularnewline
\hline 
\end{tabular}

\label{tab:score_cell_20percent}
\end{table}

\begin{table}
\caption{Coverage and length of prediction intervals for the CAR and SPDE spatial
models when holding out 80\% of entire cells of data in Minnesota.
Coverage values near 0.95 are optimal, while shorter intervals are
better.}

\begin{tabular}{|c|c|c|}
\hline 
 &
{\small{}CAR model} &
{\small{}SPDE model}\tabularnewline
\hline 
{\small{}Coverage} &
{\small{}0.981} &
{\small{}0.972}\tabularnewline
\hline 
{\small{}Mean Interval Length} &
{\small{}0.112} &
{\small{}0.103}\tabularnewline
\hline 
{\small{}Median Interval Length} &
{\small{}0.028} &
{\small{}0.022}\tabularnewline
\hline 
\end{tabular}

\label{tab:coverage_cell_20percent}
\end{table}

\paragraph{Individual tree hold-out experiment}

In Experiment 2 (individual trees held out), we have evidence (posterior
probability of 0.93) that the SPDE model is better based on the Brier
score, but the Brier score values for the two models are numerically
almost the same (Table \ref{tab:score_tree}).

\begin{table}
\caption{Predictive ability based on several predictive metric criteria for
the CAR and SPDE spatial models when holding out 5\% of trees. Smaller
values are better.}

\begin{tabular}{|c|c|c|>{\centering}p{3cm}|c|c|}
\hline 
 &
\multicolumn{3}{c|}{{\small{}Posterior mean of metric}} &
\multicolumn{2}{c|}{{\small{}Metric of posterior mean predictions}}\tabularnewline
\hline 
\hline 
 &
{\small{}CAR model} &
{\small{}SPDE model} &
{\small{}Posterior Prob. CAR < SPDE} &
{\small{}CAR model} &
{\small{}SPDE model}\tabularnewline
\hline 
{\small{}Brier} &
{\small{}0.662} &
{\small{}0.661} &
{\small{}0.07} &
{\small{}0.657} &
{\small{}0.657}\tabularnewline
\hline 
{\small{}Negative Log Density} &
{\small{}51757} &
{\small{}51626} &
{\small{}0.01} &
{\small{}50705} &
{\small{}50736}\tabularnewline
\hline 
\end{tabular}

\label{tab:score_tree}
\end{table}

\paragraph{Choice of spatial model}

The differences between models are not consistent across the various
comparisons, so there is not a clear choice. In our final data product
we use the CAR model, for three reasons. First, the CAR model has
modestly better performance when data are sparse, as is still the
case for Illinois and Indiana. Second, the model is simpler and easier
to explain, and computations can be done more quickly. Third, predictions
from the SPDE model showed boundary effects, with some taxa showing
non-negligible posterior mean values at the edges of the domain, well
away from where the taxa were present in the empirical data. This
included non-negligible values within (but near the edge of) the convex
hull of locations with data.

\section{Data product\label{sec:Data-product}}

The final data product is a dataset that contains 250 posterior samples
of the proportions of each of the 23 tree taxa at each grid cell in
the states in our domain of the northeastern United States.

For this final data product, we ran the model using the CAR specification
with all of the data (including the data held out
in the model comparison analyses) for 150,000 iterations with the
same burn-in and subsampling details as described in Section \ref{sec:Model-comparison}.
Based on graphical checks and calculation of effective sample size
values, mixing was generally reasonable, but for some of the hyperparameters
was relatively slow, particularly for less common taxa. Despite this,
mixing for the variables of substantive interest -- the proportions
-- was good, with effective sample sizes for the final product generally
near 250.

\begin{figure}
\includegraphics[scale=0.75]{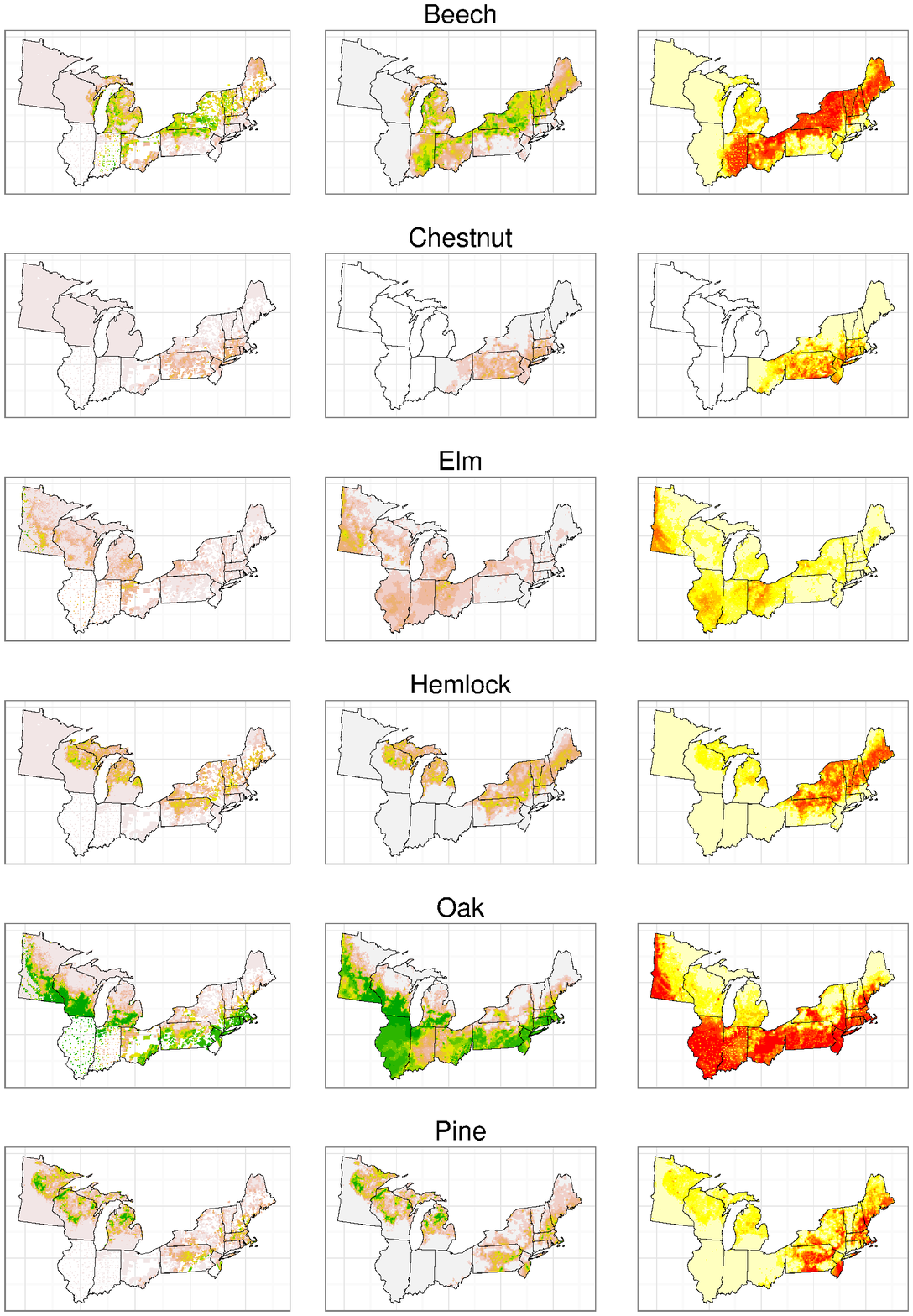}
\vspace{2mm}
\hspace{2mm}\includegraphics[scale=0.16]{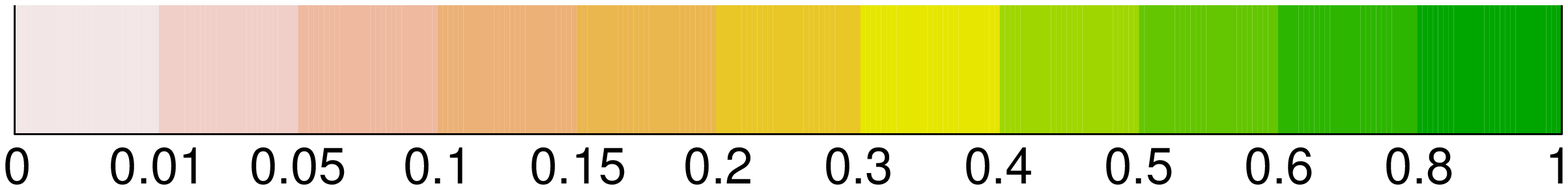}\hspace{4mm}\includegraphics[scale=0.16]{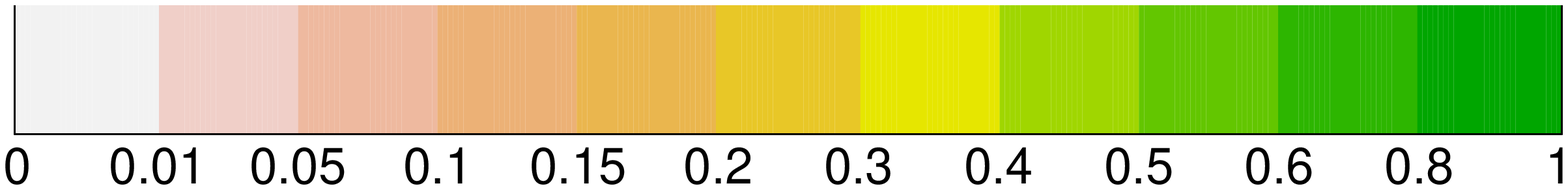}\hspace{4mm}\includegraphics[scale=0.16]{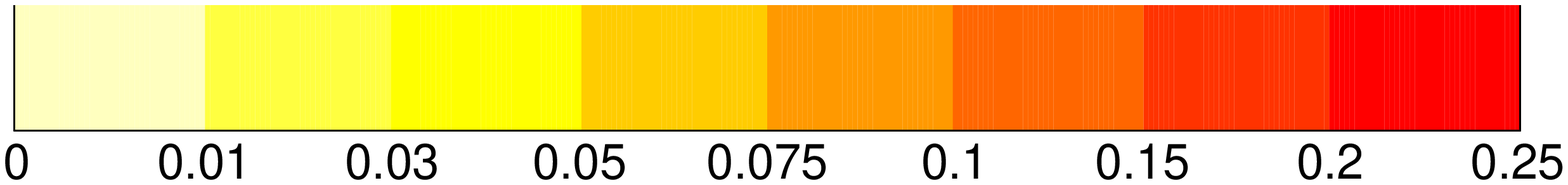}

\caption{Empirical proportions from raw data (column 1), predictions in the
form of posterior means (column 2) and uncertainty estimates in the
form of posterior standard deviations -- representing standard errors
of prediction (column 3) for select taxa. \label{fig:select_maps}}
\end{figure}

\begin{figure}
\includegraphics[scale=0.8]{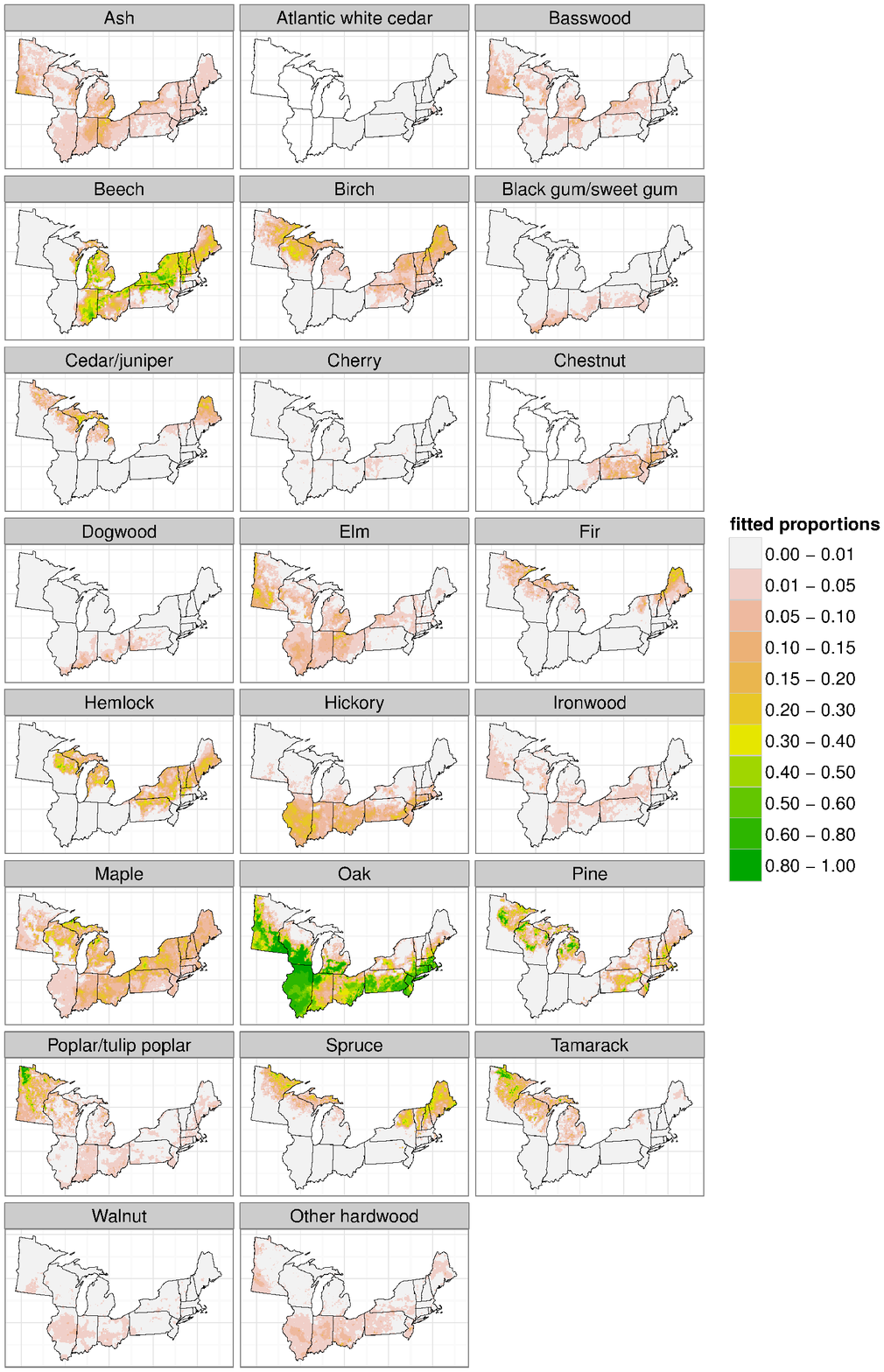}

\caption{Predictions (posterior means) for all taxa over the entire domain.\label{fig:all_predictions}}

\end{figure}

Maps of estimated composition for the full domain for several taxa
of substantive interest illustrate the results, contrasting the raw
data proportions, the posterior means, and posterior standard deviations
as pointwise estimates of uncertainty (Fig. \ref{fig:select_maps}).
We also present the posterior means for all 23 taxa (Fig. \ref{fig:all_predictions}).

The data product is publicly available at the NIS
Data Portal under the CC BY 4.0 license as version 1.0 as of January
2016 \citep{paci:etal:data:2016}. The product is in the form of a
netCDF-4 file, with dimensions x-coordinate, y-coordinate, and MCMC
iteration. There is one variable per taxon. In addition, dynamic visualizations
of the product using the Shiny tool are available at \href{https://www3.nd.edu/~paleolab/paleonproject/maps}{https://www3.nd.edu/$\sim$paleolab/paleonproject/maps}.
The PalEON Project (in particular the first author) will continue
to maintain this product, releasing new versions as additional data
in Illinois, Indiana and Ohio are digitized. Note that digitization
of data from Illinois and Indiana is ongoing, and digitization of
additional data from Ohio is planned as well. As a result, at some
point we expect to have complete data for the midwestern half of the
domain.

\section{Discussion\label{sec:Discussion}}

In the parts of the modeled region with spatially
complete data (in particular, Minnesota, Wisconsin, and Michigan),
the statistical estimates of forest composition closely match the
patterns apparent in the raw data (Fig. \ref{fig:select_maps}), as
expected.  In these areas, the estimated tree composition from the
model has the advantage of downweighting unusual or outlier values
in the empirical proportions of individual grid cells, which are likely
due to stochastic sampling variation within that grid cell (compare
the first two columns in Fig. \ref{fig:select_maps}). Some stochastic
variation is expected given that, even in the most spatially complete
regions, each grid cell contains an average of 124 trees (120-135
is typical) \citep{goring2015composition} and some cells contain
many fewer trees. Hence, some smoothing of this stochastic variation
is appropriate. This smoothing is based on information on data from
nearby cells, and the estimates from the model reflect the smooth
trends in forest composition across the spatial domain. A partial
cost is that these maps can smooth out sharp ecotones or other forms
of true spatial heterogeneity, particularly in areas with sparse data
(including areas with low tree density). For example, the sharp increase
in Elm along the Minnesota River (Fig. \ref{fig:select_maps}, first
column) likely represents a real ecotone in the settlement-era vegetation.
Vegetation gradients and ecotones were sharper in the settlement-era
forests in the upper Midwest than in contemporary forests \citep{goring2015composition},
and the modeled estimates may partially obscure this change. Users
interested in using the original unsmoothed data are directed to the
data product described earlier \citep{Gori:etal:data:2016}. Additional
investigation of other statistical representations to better capture
sharp gradients is of interest, in particular nonstationary spatial
models and use of covariates. Potential environmental covariates include
soils, firebreaks, and topography \citep{grimm1984fire,shea2014reconstructing}.
Here, however, we chose to limit our model to be exclusively a function
of spatial distance in order to avoid dependence on the environmental
drivers of pre-settlement forest composition that might lead to circular
reasoning in subsequent inferences based on this dataset. Use of covariates
could also lead to prediction that a taxa is present well beyond its
range boundary in places where data are sparse. 

A key advance of this work over prior reconstructions
of settlement-era vegetation lies in the estimates of uncertainty
across the spatial domain. These estimates of uncertainty include
the sampling uncertainty within grid cells (as do the within-grid
cell estimates of uncertainty available from the raw proportions),
but, because this is a spatial model, predictions and their associated
uncertainty estimates are also informed by the information content
of nearby cells. The maps of standard errors across species (Fig.
\ref{fig:select_maps}, third column) highlight the advantages of
this approach in areas of high data coverage (Minnesota, Wisconsin,
Michigan) and in areas of sparse coverage (e.g., Illinois, Indiana,
parts of Ohio). Where there are not large gaps in the data, the model
provides low and fairly smooth estimates of uncertainty. Uncertainty
is generally higher in the eastern subdomain than in the areas of
the midwestern subdomain with high data coverage because of missing
townships and lower sampling density even in townships with data.
In areas of sparse coverage and in areas with low tree density (e.g.,
southwestern Minnesota), the standard error of our estimates increases
appropriately. Nevetheless, these uncertainties surround reasonable
estimates of trends in composition. For example, the model does a
good job of capturing the oak ecotone in Indiana and Illinois, representing
a shift from oak savannas and woodlands to closed mesic forests (Fig.
\ref{fig:select_maps}). Experiment 1 showed that both models predicted
composition at cells with no data reasonably well, mimicking the case
of sparsely sampled data and giving confidence in the broad spatial
patterns predicted in more poorly sampled regions, particularly those
with regular, but sparse sampling that mimic the experiment (Illinois
and Indiana, but not Ohio). The apparent blockiness of uncertainty
estimates in a few places such as Ohio is caused by spatial gaps and
variations in sampling resolution. Absolute uncertainty generally
increases with abundance for all taxa (Fig. \ref{fig:select_maps},
column 3).

The exploration of alternative approaches to spatial
modeling of composition showed similar results for the SPDE and CAR
models, both in terms of prediction accuracy and performance of prediction
intervals. Small differences among the various metrics of goodness
of fit favored each model in turn, but applied users of the models
would find little pragmatic difference affecting scientific inference.
Ultimately, we slightly favor the CAR model, because it avoids the
boundary effects apparent in the SPDE model at the edges of the domain.

The models presented here estimate only the relative
abundance of tree taxa, which does not directly tell us about tree
density or other aspects of vegetation structure. This becomes a particular
limitation for interpreting vegetation where trees become sparse at
the prairie-forest transition from northern Minnesota through southern
Illinois \citep{transeau1935prairie}. Our model (correctly) predicts
that tree composition there is dominated by oak, but this is less
useful considering the sparseness of trees. This limitation can be
addressed by developing estimates of absolute abundance (e.g., biomass)
rather than compositional estimates. A gridded dataset of biomass,
stem density, and basal area is already available for Minnesota, Wisconsin,
and northern Michigan \citep{goring2015composition}, based on the
PLS data. An extension to southern Michigan, Illinois, and Indiana
is planned. We are currently developing statistical estimates of biomass
for Minnesota, Wisconsin, and Michigan using a statistical model applied
to the gridded biomass dataset, with extension to Illinois and Indiana
planned. We also plan to estimate stem density and basal area using
a similar approach to that used for biomass.

\section*{Author contributions}

CJP, ALT, and JZ developed the statistical model and code. CJP carried
out the model comparison and created the data product. CJP wrote the
paper based on an initial draft by ALT and with feedback and editing
from SJG, CVC, JWW, DJM, JAP, JZ and JSM. SJG, JAP, CVC, DJM, JSM,
and JWW led the processing and analysis of the PLS and TPS data and
assisted with interpretation of results.

\section*{Acknowledgments}

The authors are deeply indebted to all of the researchers over the
years who have preserved, collected, and digitized survey records,
in particular John Burk, Jim Dyer, Peter Marks, Robert McIntosh, Ed
Schools, Ted Sickley, Ronald Stuckey, and the Ohio Biological Survey.
We thank Madeline Ruid, Benjamin Seliger, Morgan Ripp and Daniel Handel
for processing of the southern Michigan data. Indiana and Illinois
data were made possible through the hard work of many Notre Dame undergraduates
in the McLachlan lab. This work was carried out by the PalEON Project
with support from the National Science Foundation MacroSystems Program
through grants EF-1065702, EF-1065656, DEB-1241874 and DEB-1241868
and from the Notre Dame Environmental Change Initiative. 

\bibliographystyle{/accounts/gen/vis/paciorek/latex/RSSstylefile/Chicago}

\bibliography{paperArxiv}

\section{Appendix}

\subsection{MCMC details}

Define $\bar{w}_{ip}=\frac{1}{n_{i}}{\displaystyle \sum_{j=1}^{n_{i}}W_{ijp}}$
as the average of the $W$ values for the $p$th taxon in the $i$th
grid cell and $\bar{w}_{p}=\{\bar{w}_{ip}\},$ $i=1,\ldots,m$. Let
$A$ be a diagonal matrix where $A_{ii}$ is the number of trees in
the $i$th grid cell. When there are no trees in a grid cell, $\bar{w}_{ip}=0$
and $A_{ii}=0$. For the township data, at each iteration, based on
the current values of the grid cell membership variables, $\{c_{tj}\}$,
trees are aggregated into grid cells and the calculations above can
then be carried out.

The conditional distribution for $W_{ijp}$ given the other unknowns
in the model and the data is as follows. Let $\mbox{TN}(a,b,\mu,\tau^{2})$
denote the truncated normal distribution with mean parameter $\mu$
and variance parameter $\tau^{2}$, truncated below by $a$ and above
by $b$. 

\begin{align}
W_{ijp} & \sim\begin{cases}
\mbox{TN}\big({\displaystyle \max_{p^{*}\neq y_{ij}}w_{ijp^{*}},\infty,\alpha_{y_{ij}}(s_{i}),1\big),} & \mbox{if }p\mbox{ = \ensuremath{y_{ij}}}\\
\mbox{TN}\big(-\infty,w_{ijy_{ij}},\alpha_{p}(s_{i}),1\big), & \mbox{if }p\mbox{ \ensuremath{\neq y_{ij}}}
\end{cases} \nonumber
\end{align}
In essence, the truncation value is determined by the taxon of the
$j$th tree. For a given $p$, the $W$ values for all trees in all
cells can be sampled in parallel. 

The conditional distribution of $\alpha_{p}$ is 
\begin{align}
\alpha_{p} & \sim\mbox{N}\bigg(\Big(A+Q_{p}\Big)^{-1}A\bar{w}_{p},\Big(A+Q_{p}\Big)^{-1}\bigg). \nonumber
\end{align}
where $Q_{p}=(\sigma_{p}^{2})^{-1}Q$ for the CAR model and $\left(\sigma_{p}^{2}\cdot\frac{4\pi}{\rho_{p}^{2}}\right)^{-1}Q(\rho_{p})$
for the SPDE model. For each hyperparameter, $\phi_{p}=\log\sigma_{p}$
for the CAR model and $\phi_{p}\in\{\mu_{p},(\log\sigma_{p},\log\rho_{p})\}$
for the SPDE model, we sample $\{\phi_{p},\alpha_{p}\}$ jointly,
proposing $\phi_{p}$ as a random walk and, conditional on the proposed
value of $\phi_{p}$, sampling $\alpha_{p}$ from the distribution
just above. The joint proposal is accepted or rejected as a standard
Metropolis-Hastings proposal, with adaptation of the proposal (co)variance
\citep{Shab:Well:2011}. The proposal distribution for $\phi_{p}$
is a normal distribution (bivariate for $\phi_{p}=(\log\sigma_{p},\log\rho_{p})$).

\noindent 

For the township-level data, for a given tree $j$ in township $t$,
we draw the latent tree membership variable, $c_{tj}\in\{1,\ldots,m\}$,
from a discrete distribution by normalizing posterior weights, $\{\psi_{1}L_{tj1},\ldots,\psi_{m}L_{tjm}\}$,
produced by multiplying the prior weights by a likelihood contribution,
$L_{tji}$, $i=1,\ldots,m$. $L_{tji}$ is the density of the latent
$W_{tj1},\ldots,W_{tjP}$ values for the given tree under the condition
that $c_{tj}=i$, namely the product of independent normal densities,
$W_{tjp}\sim\mbox{N}(\alpha_{p}(s_{i}),1)$, over $p=1,\ldots,P$.
Thus the posterior reweights the prior based on how consistent the
current $W_{tj}$ values for a tree are with the $\alpha$ values
for the candidate grid cells.

\subsection{Estimating $\theta_{p}(s)$ via Monte Carlo integration}

In the latent variable representation, $\theta_{p}(s)$ never appears
explicitly and cannot be calculated in closed form. Instead we use
Monte Carlo integration over $W_{ijp},\, p=1,\ldots,P$ to estimate
$\theta_{p}(s_{i})$. The quantity $\theta_{p}(s_{i})=\mbox{Prob}(W_{ijp}={\displaystyle \max_{p^{*}}W_{ijp^{*}})}$
defines the probability of taxon $p$ at grid cell $i$. This requires
one to choose the number of Monte Carlo samples, which we set at 10000,
effectively sampling 10000 hypothetical trees and estimating the probabilities
of the different taxa in the population from the empirical proportions
in this sample of trees. For each of the saved MCMC samples, $k=1,\ldots K$,
we estimate $\theta_{p}^{(k)}(s_{i})$ numerically. Specifically,
for $t=1,\ldots,10000$ samples (i.e., hypothetical trees), we independently
draw
\[
W_{itp}^{(k)}\sim\mbox{N}(\alpha_{p}^{(k)}(s_{i}),1),\, p=1,\ldots,P
\]
and estimate using 
\[
\theta_{p}^{(k)}(s_{i})\approx\frac{1}{10000}{\displaystyle \sum_{t=1}^{10000}1(W_{itp}^{(k)}={\displaystyle \max_{p^{*}}W_{itp^{*}}^{(k)})}},\, p=1,\ldots,P
\]
where $1(\cdot)$ is the indicator function that evaluates to 1 if
the expression is true and 0 if false. In other words, we calculate
the proportion of times that the maximum of $W_{itp},\, p=1,\ldots,P$
corresponds to taxon $p$. Considering $\theta_{p}^{(k)}(s_{i}),\, k=1,\ldots,K$,
we have a sample from the posterior of $\theta_{p}(s_{i})$.

\end{document}